# Modulating Optical Properties through Cation Substitution: Composition-Property Relationships in $M^I_3M^{III}P_3O_9N:Eu^{2+}$ ($M^I$=Na, K; $M^{III}$=Al, Ga, In)

Nakyung Lee[1,2], Justyna Zeler[3], Małgorzata Sójka[1,2], Eugeniusz Zych[3], Jakoah Brgoch[1,2]

[1]*Department of Chemistry, University of Houston, Houston, Texas 77204, USA*
[2]*Texas Center for Superconductivity, University of Houston, Houston, Texas 77204, USA*
[3]*Faculty of Chemistry, University of Wroclaw, 14. F Joliot Curie Street, 50-383 Wroclaw, Poland*

**Abstract**

Developing phosphors with narrow photoluminescence emission peaks and high chromatic stability holds significant importance in light-emitting diode (LED) display technologies, where a wide color gamut is essential to achieve the Rec. 2020 specifications. This research focuses on the optical properties of a solid solution: $M^I_{2.97}Eu_{0.015}M^{III}P_3O_9N$ [$M^I$=Na, K; $M^{III}$=Al, (Al$_{0.75}$Ga$_{0.25}$), (Al$_{0.5}$Ga$_{0.5}$), (Al$_{0.25}$Ga$_{0.75}$), Ga, (Ga$_{0.75}$In$_{0.25}$), (Ga$_{0.5}$In$_{0.5}$)] to understand how the narrow-emitting photoluminescence in $K_3AlP_3O_9N:Eu^{2+}$ can evolve during host structure cation substitution. Photoluminescence measurements at low temperature (15 K) support that $Eu^{2+}$ replaces three crystallographically independent $Na^+$ sites in $Na_{2.97}Eu_{0.015}AlP_3O_9N$, similar to the parent $K^+$ phosphor, but substituting $Ga^{3+}$ and $In^{3+}$ for $Al^{3+}$ leads to a change in $Eu^{2+}$ site preference, narrowing the full-width-at-half-maximum (*fwhm*) of the emission peak. The chromatic stability and photoluminescence quantum yield are also enhanced with higher $Ga^{3+}$ content in the host but not with $In^{3+}$. Thermoluminescence analysis indicates the relationship between trap states and the enhanced quantum yield with $Ga^{3+}$ leads to the series' best performance. The analysis of the $M^I_{2.97}Eu_{0.015}M^{III}P_3O_9N$ series offers insight into the potential method for modulating optical properties with cation substitution in the host structure.

## 1. Introduction

Pursuing sustainable alternatives to traditional lighting and more portable display technologies has led to the rapid development of phosphor-converted light-emitting diodes (pc-LEDs). Their energy efficiency and prolonged operating lifespan guarantee a reduction in electricity consumption and promise a substantial decrease in their environmental footprint.[1,2] Pc-LEDs generate white light by coating a blue LED ($\lambda_{ex}$=450 nm) with yellow and red phosphors, which absorb the blue emission and down-shift it into a broad spectrum.[3,4] Alternatively, a combination of near-ultraviolet (n-UV) LED chips ($\lambda_{ex}$=380 nm-420 nm) and blue, green, and red-emitting phosphors is another proposed strategy capable of generating higher-quality white light by covering a more significant portion of the visible spectrum. An added advantage of this approach is regulating the amount of blue LED light produced by the device in what is being called "human-centric" lighting.[5,6] Avoiding excessive exposure to blue light, which suppresses melatonin production and stimulates the brain, and replicating color temperature shifts throughout the day is valuable for regulating the human circadian rhythm. LED lighting can address these issues, provided phosphor conversion materials can be developed.

The U.S. Department of Energy has previously outlined the criteria for developing commercial-ready phosphors capable of generating white light.[7] Phosphors must be capable of absorbing and efficiently down-shifted light from blue or violet LEDs and produce narrow photoluminescence emission with a full-width-at-half-maximum (*fwhm*) that is ≈30 nm (red=794 cm$^{-1}$, green=1148 cm$^{-1}$, blue=1485 cm$^{-1}$). The narrow emission is vital to enhance lighting efficacy by minimizing the emission of photons in the near-IR and UV regions. It is also crucial for display applications to generate a wider color gamut and achieve the Rec. 2020 specifications.[8–10] Commercially relevant phosphors should also withstand average LED operating temperatures of 150°C without succumbing to thermally induced changes in the optical properties. For instance, a phosphor's emission intensity should change by less than 50% (ideally less than 4%) from room temperature up to 150°C to overcome thermal quenching.[11] The phosphor's emission color must also not vary as a function of temperature shifting by less than a 3-step MacAdam ellipse.[12]

Given such specific requirements, new materials are constantly being tested. (Oxy-)nitrides substituted with rare-earth (RE) activator ions, like $Eu^{2+}$, are one major class of phosphors studied today motivated by their propensity to strongly absorb violet/blue light and re-emit these photons across the broad visible spectrum via the parity-allowed $4f^7 \leftrightarrow 4f^65d^1$ electronic transitions.[8,13,14] The energy position of these $Eu^{2+}$ transitions is set based on the nature of the host structure's coordinating atoms and the geometry of the crystallographic substitution site. The resulting nephelauxetic effect stabilizes (lowers) the rare-earth $5d$ orbitals in energy by reducing the inter-electron repulsion of activator ions.[15–17] This effect depends mainly on the anion type, with $N^{3-}$ causing a more considerable decrease in energy than $O^{2-}$, while $O^{2-}$ is more effective than $F^-$. The $5d$ orbitals of the lanthanide ion are further separated by crystal field splitting (CFS) effects, which depend primarily on the coordination number, volume of the cation site, and symmetry of the site.[18,19] The combination of effects is exemplified by comparing the phosphorus oxynitride $Ba_2PO_3N:Eu^{2+}$ with the $[BaO_7N_3]$ and $[BaO_7N_2]$ sites generating cyan emission ($\lambda_{em}$=530 nm), whereas the phosphate $Ba_2P_2O_7:Eu^{2+}$ shows blue-shifted emission ($\lambda_{em}$=420 nm) from the $[BaO_{10}]$ and $[BaO_7]$ sites.[20,21]

The position of the electronic transitions and associated emission color is not necessarily binary. Forming a solid solution through meticulous elemental substitution is one approach that enables tuning of the phosphors' photoluminescent properties.[2,22,23] These substitutions are valuable on either the anion or the cation sublattice. The most common is varying the cation composition to change the properties gradually. For example, the $Rb_2(Ca_{1-x}Sr_x)P_2O_7:Eu^{2+}$ system replaces the larger cation $Sr^{2+}$($r_{6\text{-coord}}$=1.18 Å) for the smaller $Ca^{2+}$($r_{6\text{-coord}}$=1.00 Å) expanding the unit-cell, generating a continuous blue-shifted emission from 612 nm to 567 nm.[24] Similarly, oxynitrides can also benefit from forming solid solutions, provided cation charge compensation mechanisms are available. The oxynitridosilicate $Sr_{2-x}La_xSiO_{4-x}N_x$ solid solution incorporates nitrogen with the cation charge balanced out by adding $La^{3+}$ for $Sr^{2+}$, resulting in a tunable red-shifted emission from 550 nm to 700 nm with increasing $N^{3-}$ content.[25] Varying the composition does not just shift the maximum of the emission peak, but it can holistically impact the optical properties. In the $La_5(Si_{2+x}B_{1-x})(O_{13-x}N_x):Ce^{3+}$ system, the increase of $N^{3-}$ shifts the emission from violet ($\lambda_{em}$=421 nm) to blue ($\lambda_{em}$=463 nm) region while simultaneously enhancing the thermal quenching temperature ($T_{50}$) from 120°C ($x$=0) to 202°C ($x$=0.7).[22] Solid solution formation can also create more complex outcomes like changing rare-earth substitution site preference in the phosphor. The $(Rb_{1-x}K_x)_2CaPO_4F:Eu^{2+}$ solid solution shows that decreasing the $Rb^+$ site volume while expanding the $Ca^{2+}$ site by incorporating $K^+$ can cause significant changes by varying substitution site preference.[26] The pure $Rb^+$ composition shows strong site preference on $[Rb(2)O_8F_2]$ and no emission from $[CaO_4F_2]$, whereas RbKCaPO$_4$F, $Eu^{2+}$ shows a stronger emission peak from $[Rb(1)O_8F_2]$. Finally, a higher $K^+$ ratio than $Rb^+$ pushes the emission red-shifted by letting $Eu^{2+}$ mostly occupy $[CaO_4F_2]$ sites than other $Rb^+$ sites. The study of solid solutions plays a crucial role in controlling the optical properties of phosphors.

One oxynitride discovered by our group where solid solutions may prove interesting for tuning the optical properties is the $K_3AlP_3O_9N:Eu^{2+}$ system.[27] This material's narrow emission (*fwhm*=45 nm, 2,110 cm$^{-1}$) and reasonable photoluminescent quantum yield (PLQY) of 60(4)%, holding promise in violet excited human-centric display concepts. Based on this previous study, the complete cation substitution of K$^+$ with a smaller Na$^+$ is tested to tune the emission color and investigate the influence of $M^I$ site substitution on the optical properties. Subsequent investigations into solid solution substitutions on the $M^{III}$ site following Na$_{2.97}$Eu$_{0.015}$$M^{III}$P$_3$O$_9$N [$M^{III}$=Al, (Al$_{0.75}$Ga$_{0.25}$), (Al$_{0.5}$Ga$_{0.5}$), (Al$_{0.25}$Ga$_{0.75}$), Ga, (Ga$_{0.75}$In$_{0.25}$), (Ga$_{0.5}$In$_{0.5}$)] are performed to understand the chemistry of these Na$^+$ analogs in depth. The properties are investigated using low-temperature emission, temperature-dependent emission, and thermoluminescence (TL) glow curve measurements. The insights gained from these results deepen our understanding of how elemental substitution can influence optical properties. In turn, these findings may facilitate deliberate development of new phosphors for pc-LEDs and careful tuning of their properties.

## 2. Experimental Procedure

### 2.1 Materials Synthesis

The solid solutions of (Na$_{1-2x}$Eu$_x$)$_3$AlP$_3$O$_9$N ($x$=0.001, 0.0025, 0.005, 0.01, 0.02) and Na$_{2.97}$Eu$_{0.015}$$M^{III}$P$_3$O$_9$N [$M^{III}$=Al, (Al$_{0.75}$Ga$_{0.25}$), (Al$_{0.5}$Ga$_{0.5}$), (Al$_{0.25}$Ga$_{0.75}$), Ga, (Ga$_{0.75}$In$_{0.25}$), (Ga$_{0.5}$In$_{0.5}$)] were synthesized by solid-state reactions starting from KPO$_3$, Na$_6$[(PO$_3$)$_6$] (Sigma-Aldrich, 96%), Al$_2$O$_3$ (Alfa Aesar, 99%), Ga$_2$O$_3$ (Alfa Aesar, 99.99%), In$_2$O$_3$ (Alfa Aesar, 99.9%), and Eu$_2$O$_3$ (Alfa Aesar, 99.99%). KPO$_3$ was prepared by dehydrating KH$_2$PO$_4$ (Alfa Aesar, 99.0%) at 350 °C for 12 h in air. Each component was weighed out in the appropriate stoichiometric ratio, and the starting reagents were mixed and ground in an agate mortar and pestle using acetone as a wetting medium. Powders were further milled for 30 min in a high-energy ball mill (Spex 800 M Mixer/Mill) using polystyrene vials. The mixtures were then pressed into a 6 mm diameter pellet and placed on a bed of sacrificial powder in a boron nitride crucible. The pellets were heated at 800°C for 12 h with a heating and cooling rate of 3°C/min under flowing NH$_3$ gas.

### 2.2 Structural and Optical Characterization

Powder X-ray diffractograms of all products were collected using an X'Pert PANalytical Empyrean 3 equipped with Cu K$\alpha$ ($\lambda$ = 1.5406 Å) as the radiation source. Le Bail refinements were performed using the General Structural Analysis System II (GSAS-II) software.[28] The background was described using a Chebyshev-1 function, and the peak shapes were modeled by a pseudo-Voigt function. Scanning electron microscopy (SEM) micrographs and energy-dispersive X-ray spectroscopy (EDS) elemental mappings were collected using a Phenom Pharos scanning electron microscope attached with a 25 mm$^2$ silicon drift detector energy-dispersive X-ray spectrometer (Thermo Fisher Scientific). An accelerating voltage of 15 kV and an emission current of 12 μA were used.

The bandgap of the Na$_3$$M^{III}$P$_3$O$_9$N host series was calculated using density functional theory (DFT) calculations with the Vienna ab initio simulation (VASP) package within the projector-augmented wave (PAW) method.[29,30] The generalized gradient approximation Perdew-Burke-Ernzerhof (PBE) functional was used for the energy calculations with a plane wave energy cutoff of 520 eV and the Brillouin zones were integrated with a $k$-point density grid of at least 100/Å$^{-3}$ (reciprocal lattice volume).[31]

Steady-state photoluminescence and quantum yield measurements involved mixing the polycrystalline products in an optically transparent silicon epoxy (United Adhesives Inc., OP 4036) and depositing the combination onto a quartz slide (Chemglass). The excitation and emission spectra were obtained using a PTI fluorescence spectrophotometer equipped with a 75 W xenon arc lamp excitation source. The PLQY was determined following the method of de Mello *et al.*[32] using a Spectralon-coated integrating sphere (150 mm diameter, Labsphere) with an excitation wavelength of 340 nm. Temperature-dependent photoluminescence was recorded between 15 K to 600 K using a FLS1000 Fluorescence Spectrometer from



Edinburgh Instruments Ltd. The samples were mounted on a copper holder in a closed-cycle helium cryostat from Lake Shore Cryotronics, using Silver Adhesive 503 glue from Electron Microscopy Sciences. A 450 W continuous xenon arc lamp was used as the excitation source for these measurements. A TMS302-X double grating excitation and emission monochromators with 2x325 mm focal lengths were used, and the luminescence signal was recorded with a Hamamatsu R928P high-gain photomultiplier detector thermoelectrically cooled to −22°C. The excitation spectra were corrected for the incident light intensity, and the emission ones were corrected for the emission channel spectral sensitivity.

Thermoluminescence (TL) experiments were performed using Lexsyg Research Fully Automated TL/OSL Reader from Freiberg Instruments GmbH. An X-ray lamp VF-50J RTG with W-anode operated under 12 kV and 0.1 mA was used as a charging source. The phosphor samples' irradiation with X-rays (12 kV, 0.1 mA) was performed for 10 seconds and then TL glow curves were recorded using a 9235QB-type photomultiplier from ET Enterprises in the range from 303 K to 773 K. All experiments were controlled through LexStudio 2 software.

## 3. Results and Discussion

### 3.1 Distinctive Optical Behavior in $M^I_{2.97}Eu_{0.015}AlP_3O_9N$ ($M^I$=Na, K)

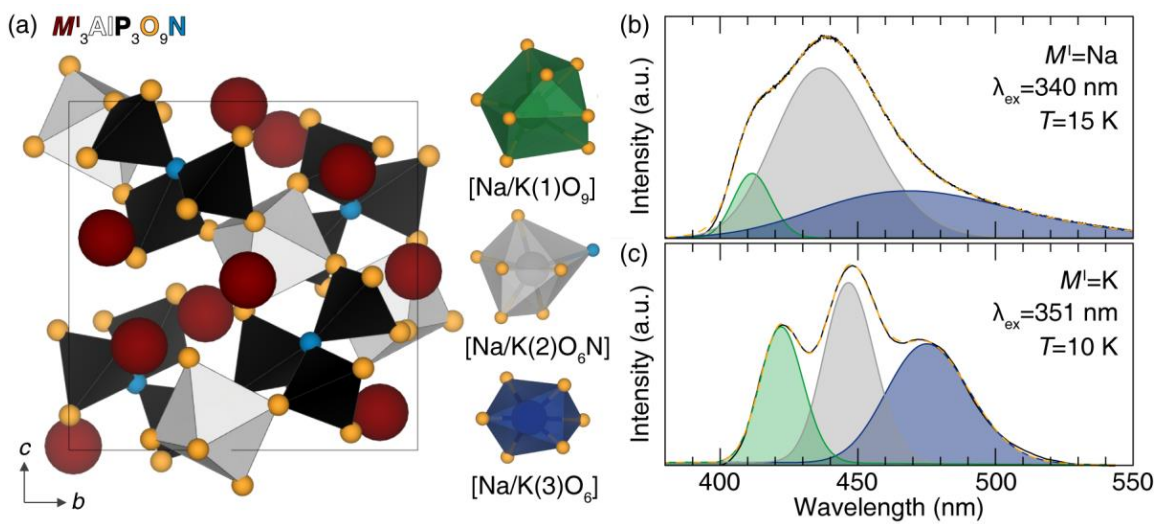

**Figure 1.** (a) Crystal structure of $M^I_3AlP_3O_9N$ (ICSD #78373) and three crystallographically independent Na$^+$/K$^+$ sites for Eu$^{2+}$ substitution. A bond length of < 2.96 Å creates a [$M^I(1)O_6$] polyhedron that is severely elongated and distorted. However, when the bond length is ≈2.98 Å, a 9-coordinated polyhedron [$M^I(1)O_9$] creates a more reasonable site geometry in the unit cell that is more coherent with the photoluminescent characteristics. Photoluminescence deconvoluted emission spectrum of (b) Na$_{2.97}$Eu$_{0.015}$AlP$_3$O$_9$N and (c) K$_{2.97}$Eu$_{0.015}$AlP$_3$O$_9$N at low temperatures. Gaussian deconvolution was made by first converting the spectra into energy scale by means of Jacobian transformation.[33]

The previously reported phosphor, K$_3$AlP$_3$O$_9$N:Eu$^{2+}$, crystallizes in cubic space group $P2_13$ (No. 198), and is part of a broader isostructural series following $M^I_3M^{III}P_3O_9N$ ($M^I$=Na; K, $M^{III}$=Al; Ga; Cr; Fe; Mn).[34] Illustrated in **Figure 1a**, the crystal structure was originally solved by single-crystal X-ray diffraction with three [PO$_3$N] tetrahedra connected through a nitrogen-occupied vertex that creates a [N(PO$_3$)$_3$] unit. The [AlO$_6$] octahedra share an O atom from six different [PO$_3$N] tetrahedra, producing an Al[N(PO$_3$)$_3$] backbone. There are three crystallographically independent $M^I$ sites in the crystal structure: [$M^I(1)O_9$], [$M^I(2)O_6N$], and [$M^I(3)O_6$]. Although the [$M^I(1)O_9$] site was reported as a six-coordinated site in the original reference, a closer re-analysis suggests a nine-coordinated environment as a better description based on the bond



lengths of the adjacent atoms and space-filling in the unit cell.[27]

The previous $K_3AlP_3O_9N$:$Eu^{2+}$ study showed a 454 nm (blue) emission at a room temperature with bandwidth of 45 nm (2,110 cm$^{-1}$) under 400 nm excitation. The relatively narrow photoluminescence stemmed from two factors: site-selective excitation ($\lambda_{ex}$=405 nm) efficiently removes the emission from [K(1)O$_9$], and a preferential thermal quenching of the [K(3)O$_6$] site at room temperature. As a result, only $Eu^{2+}$ on the [K(2)O$_6$N] site contributes to the emission under violet excitation. Building on this prior investigation, this work first explores the substitution of $K^+$ with the smaller $Na^+$ ions ($K^+$: $r_{6\text{-coord}}$ = 1.38 Å, $r_{7\text{-coord}}$ = 1.46 Å, $r_{9\text{-coord}}$ = 1.55 Å; $Na^+$: $r_{6\text{-coord}}$ = 1.02 Å, $r_{7\text{-coord}}$ = 1.12 Å, $r_{9\text{-coord}}$ = 1.24 Å)[35] to understand the relationship between composition and the effects of cation substitution while tuning the emission color.

The reported synthesis of $Na_3AlP_3O_9N$ used a synthesis temperature of 800°C with AlN in a molten flow of sodium polyphosphates[36,37]. In the study here, polycrystalline samples of $(Na_{1-2x}Eu_x)_3AlP_3O_9N$ (x=0.001, 0.0025, 0.005, 0.01, 0.02) were obtained at 800°C under flowing NH$_3$, which is used for supplying nitrogen in the target compositions and for reducing $Eu^{3+}$ to $Eu^{2+}$. The products had no notable impurities, and their peaks matched the reference pattern, as confirmed by powder X-ray diffraction (**Figure S1a**). Analyzing the unit-cell dimensions using Le Bail refinements shows that the lattice parameters increase with the substitution of the larger $Eu^{2+}$ ion for the smaller $Na^+$ ion ($Eu^{2+}$ $r_{6\text{-coord}}$=1.17 Å, $r_{7\text{-coord}}$=1.2 Å, $r_{9\text{-coord}}$=1.3 Å) (**Figure S1b**). The refined lattice parameters are provided in **Table S1**.

Substituting $Eu^{2+}$ into the solid solution $(Na_{1-2x}Eu_x)_3AlP_3O_9N$ generates deep blue luminescence with an excitation band ranging from 280 nm to 400 nm (see **Figure S1c**). The excitation spectra for x=0.001, 0.0025, and 0.005 shows little change in shape; however, a distinct shoulder appears around 375 nm in the excitation spectrum for x=0.01 and 0.02. A similar trend is observed when collecting the emission spectrum using $\lambda_{ex}$ = 340 nm, with a shoulder extending from 450 nm to 550 nm. As a result, there is a slight red shift in the deep blue emission, with $\lambda_{em,max}$=427 nm at x=0.001 shifting to $\lambda_{em,max}$=432 nm at x=0.02 as the concentration of $Eu^{2+}$ increases. Generally, a higher concentration of larger $Eu^{2+}$ ions expands the unit cell and the volume of the substituted sites, leading to a blue shift in the $\lambda_{em,max}$ position accompanied by lower crystal field splitting (CFS). However, in this case, the presence of shoulders in both photoluminescent excitation and emission spectra suggests that $Eu^{2+}$ prefers to substitute onto [Na(2)O$_6$N] or [Na(3)O$_6$], resulting in a longer emission peak wavelength at higher rare earth concentrations.

The optimal $Eu^{2+}$ concentration in the $(Na_{1-2x}Eu_x)_3AlP_3O_9N$ series was determined by measuring the PLQY under 340 nm excitation (**Figure S1d**). At x=0.005, PLQY reaches the highest value among the series with 21(2) %. Beyond x=0.01, the PLQY drops rapidly due to the increased probability of the energy transfer between $Eu^{2+}$ ions. In higher $Eu^{2+}$ concentrations, there is less separation between $Eu^{2+}$ ions, increasing the likelihood of energy transfer and enabling electrons to find quenching sites, inducing a non-radiative relaxation. The critical distance of energy migration, $R_c$, between two activators can be calculated with **Equation 1**,

$$R_c = 2\left(\frac{3V}{4\pi x_c n}\right)^{1/3} \quad (1)$$

where $V$ is the volume of the unit cell, $x_c$ is the critical concentration of the activator at maximum emission intensity, and $n$ is the sum of the Wyckoff positions of the crystallographically independent substitution sites.[38] The calculated $R_c$ here is 26.2 Å. The long-distance non-radiative energy transfer generally occurs through electron exchange interaction or electric multipolar interaction. To decide what type of interaction among dipole-dipole (d-d), dipole-quaternary (d-q), and quaternary-quaternary (q-q) interactions causes the



quenching, the emission intensity (*I*) at each activator concentration is analyzed with **Equation 2**,

$$\frac{I}{x} = k[1 + \beta(x)^{\theta/3}]^{-1} \tag{2}$$

where *x* is the activator concentration above the critical concentration, *k* and *β* are constants of a given host, and *θ* is the electric multipolar character.[39] The energy transfer mode of electric multipolar interaction can be confirmed by the values of *θ*=6, 8, and 10, which correspond to d-d, d-q, and q-q interaction, respectively. With an assumption of $\beta(x)^{\theta/3} \gg 1$, **Equation 2** can be simplified as a linear equation with slope $-\theta/3$.

$$\log I/x = k' - \theta/3 \log x \quad (k' = \log k - \log \beta) \tag{3}$$

The plot of $\log I/x$ versus $\log x$ of $Na_{3-2x}Eu_xP_3O_9N$ shows the slope of −2.0(2), indicating *θ*=6 (**Figure S2**). Thus, the electric dipole-dipole interaction is the main source of non-radiative concentration quenching, producing the low PLQY in this system at high $Eu^{2+}$ concentrations.

Additional differences in the photoluminescent properties of $K_3AlP_3O_9N:Eu^{2+}$ and $Na_3AlP_3O_9N:Eu^{2+}$ are immediately apparent at low temperatures. The emission of the $Na^+$ analog is blue-shifted and broader ($\lambda_{em}$ = 429 nm, *fwhm* = 58 nm, 3,217 cm$^{-1}$) compared to the reported $K_3AlP_3O_9N:Eu^{2+}$ ($\lambda_{em}$ = 445 nm, *fwhm* = 54 nm, 2,865 cm$^{-1}$) under 340 nm excitation and at low temperature (77 K). Deconvoluting the spectra into their components indicates that $Na_3AlP_3O_9N:Eu^{2+}$ (**Figure 1b**) and $K_3AlP_3O_9N:Eu^{2+}$ (**Figure 1c**) can both be fit by three Gaussian peaks with the peak center arising from $Eu^{2+}$ on the $[M^I(1)O_9]$ position shifting from 416 nm to 410 nm, $Eu^{2+}$ on the $[M^I(2)O_6N]$ position moving from 447 nm to 432nm, and $Eu^{2+}$ on the $[M^I(3)O_6]$ position moving from 465 nm to 462 nm, for the $K^+$ and $Na^+$ analogs, respectively. The observed blue shift of each peak is somewhat of a surprise considering the unit cell volume decreased from 905.5 Å$^3$ to 801.8 Å$^3$ when fully substituting $K^+$ with $Na^+$. This would normally suggest an expected redshift of the emission spectrum to the cyan or even green region of the visible spectrum due to stronger CFS. This unexpected blue shift, therefore, must be attributed to weaker CFS stemming from local distortions as the much larger $Eu^{2+}$ is stuffed on the smaller $Na^+$ site. Such changes are not uncommon. In fact, the most popular example is the blue shift between the commercial yellow-emitting phosphor $Y_3Al_5O_{12}:Ce^{3+}$ and its derivative $Lu_3Al_5O_{12}:Ce^{3+}$. Despite replacing the larger $Y^{3+}$ with the smaller $Lu^{3+}$, $Lu_3Al_5O_{12}:Ce^{3+}$ shows a blue-shifted green emission because of local distortion coming from the dodecahedral $[LuO_8]$ site than $[YO_8]$ and generates smaller crystal field splitting ($\varepsilon_{cfs,YAG}$=3.34 eV, $\varepsilon_{cfs,LuAG}$=3.12 eV).[40,41]



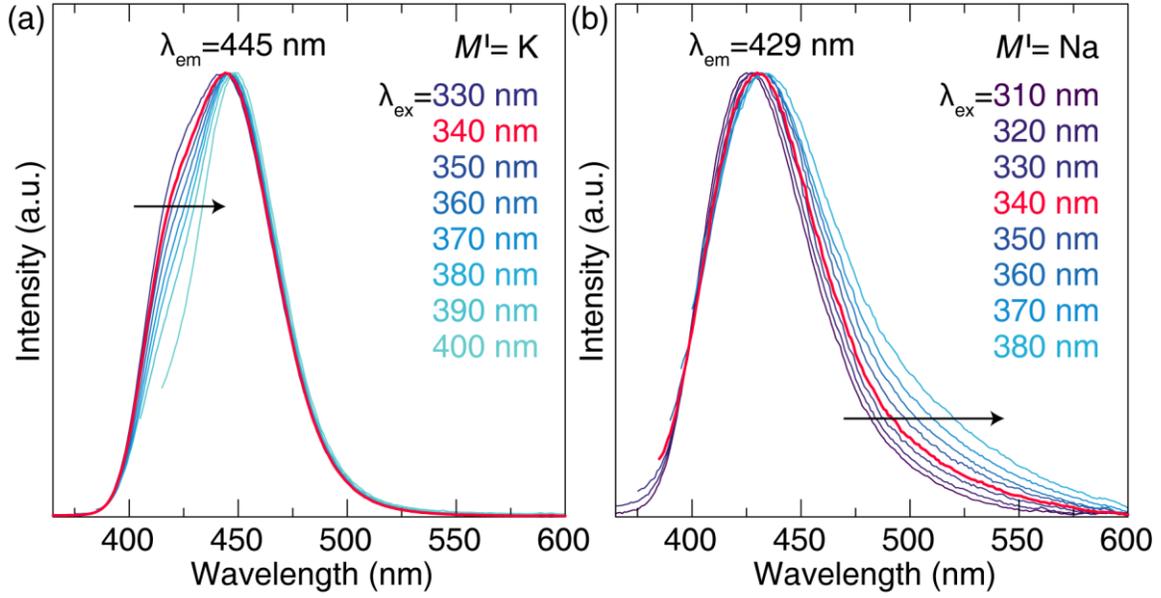

**Figure 2.** Emission spectra of (a) $K_3AlP_3O_9N:Eu^{2+}$ and (b) $Na_{2.97}Eu_{0.015}AlP_3O_9N$ under different excitation wavelengths recorded at room temperature show that site-selective excitation on [Na(1)O_9] and preferential quenching on [Na(3)O_6] do not happen in $M^I$=Na unlike $M^I$=K.

Further differences in the optical properties are revealed by monitoring changes in the emission spectrum under various excitation wavelengths (**Figure 2**). On the high-energy side of the emission spectrum, the emission from [Na(1)O_9] remains unchanged with respect to excitation wavelength, unlike the [K(1)O_9] site, which exhibits a significant narrowing of the emission peak. Conversely, longer excitation wavelengths promote longer-wavelength emission from [Na(3)O_6], whereas this phenomenon is not observed for [K(3)O_6]. The result is that site-selective excitation and preferential thermal quenching fail to narrow the $M^I$ = Na analog. In fact, the emission broadens with excitation, in contrast to the narrower emission obtained for $M^I$ = K with longer excitation wavelengths. The excitation spectra of $M^I$=Na and $M^I$=K for $\lambda_{em}$ = 420 nm and $\lambda_{em}$ = 450 nm were also measured (**Figure S3**). In both cases, longer wavelength emissions arise from longer wavelength excitations. The shift in excitation of $M^I$=Na is less than $M^I$=K, moving from 329 nm for $\lambda_{em}$ = 420 nm to 340 nm for $\lambda_{em}$ = 450 nm, while $M^I$=K moves from 346 nm to 352 nm.

**3.2 Fine-tuning the Emission Spectrum through Solid-Solutioning: $Na_{2.97}Eu_{0.015}M^{III}P_3O_9N$ ($M^{III}$=Al, Ga, In)**



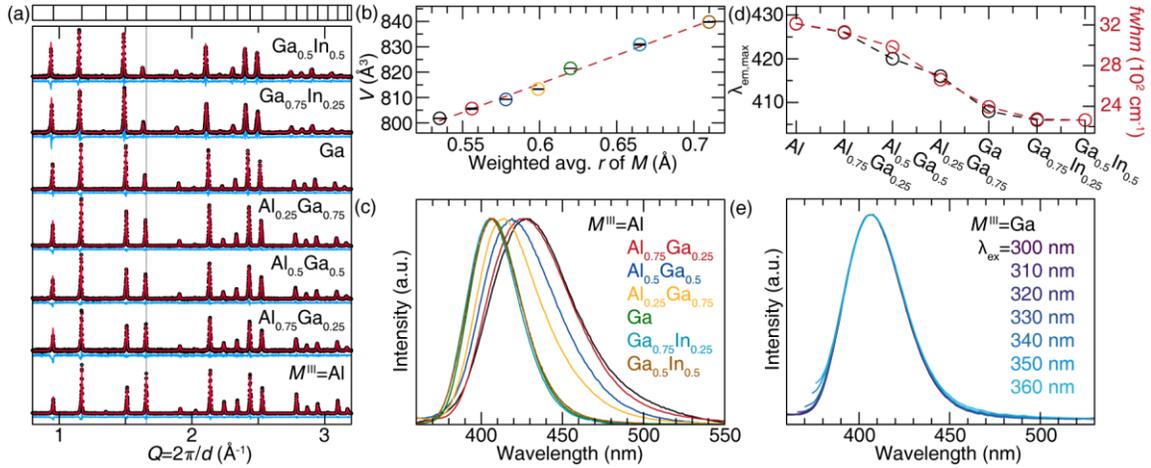

**Figure 3.** (a) Le Bail refinements of $Na_{2.97}Eu_{0.015}M^{III}P_3O_9N$ ($M^{III}$=Al, Ga, In) starting from $Na_3AlP_3O_9N$ structure (ICSD #78373). Peak positions shift towards smaller values in $Q$ space along with increments of larger $M^{III}$ elements, $Ga^{3+}$ than $Al^{3+}$ and $In^{3+}$ than $Ga^{3+}$, which are emphasized by a greyline around 1.6 Å$^{-1}$. (b) Calculated volumes of the entire series demonstrate a linear increase according to the weighted average radius of $M^{III}$. (c) Normalized emission spectra of $Na_{2.97}Eu_{0.015}M^{III}P_3O_9N$ under $\lambda_{ex}$=340 nm. (d) The peak positions and *fwhm* of these emission spectra show that the increment of $Ga^{3+}$ induces more blue-shifted and narrower emissions. However, the substitution of $Ga^{3+}$ by $In^{3+}$ barely influences the emission properties. (e) Emission spectra under different excitations of $M^{III}$=Ga reveal no significant changes compared to $M^I$=K and $M^I$=Na.

Substitution on the $M^I$ site clearly results in significant changes to the optical properties. Therefore, the full extent of these changes was further probed by exchanging the $M^{III}$ site. Synthesizing a series of compositional derivatives following $Na_{2.97}Eu_{0.015}M^{III}P_3O_9N$ [$M^{III}$=Al, (Al$_{0.75}$Ga$_{0.25}$), (Al$_{0.5}$Ga$_{0.5}$), (Al$_{0.25}$Ga$_{0.75}$), Ga, (Ga$_{0.75}$In$_{0.25}$), (Ga$_{0.5}$In$_{0.5}$)] was possible under the same synthetic conditions. The purity of polycrystalline products was confirmed by powder X-ray diffraction (**Figure 3a**). The peak positions shift to smaller 2θ when $Al^{3+}$ is substituted by $Ga^{3+}$ and $Ga^{3+}$ is substituted by $In^{3+}$, indicating the larger $M^{III}$ ion replaces the smaller ion ($Al^{3+}r_{6-coord}$=0.535 Å, $Ga^{3+}r_{6-coord}$=0.62 Å, $In^{3+}r_{6-coord}$=0.8 Å).[35] All attempts to extend the $Na_{2.97}Eu_{0.015}M^{III}P_3O_9N$ series increasing the $In^{3+}$ content to $M^{III}$=Ga$_{0.25}$In$_{0.75}$ and $M^{III}$=In led to impurities. Indeed, the diffractogram of $M^{III}$=Ga$_{0.25}$In$_{0.75}$ (**Figure S4a**) includes unmatched impurity peaks while analysis by SEM suggests that the product possibly contained indium metal as the ammonia reduced $In_2O_3$ (**Figure S4b**). In the case of $M^{III}$=In, the indium metal particles were large enough to be observed by eye.

Le Bail refinements (**Table S2**) of the $Na_{2.97}Eu_{0.015}M^{III}P_3O_9N$ series were performed using the $Na_3AlP_3O_9N$ crystal structure file (ICSD #78373) as the starting point. The calculated unit cell volumes of the solid solution series show a linear expansion along the weighted average radius of $M^{III}$, supporting that $Ga^{3+}$ and $In^{3+}$ replaced $Al^{3+}$ (**Figure 3b**). Micrographs collected on an SEM and a subsequent EDS analysis of $Na_{2.97}Eu_{0.015}AlP_3O_9N$ and $Na_{2.97}Eu_{0.015}GaP_3O_9N$ (**Figure S5**) also show that all elements are uniformly distributed in the examined particles and there is no apparent phase or elemental separation. Since the $Eu^{2+}$ concentration (~0.09%) is below the instrument's detection limit, the EDS result for $Eu^{2+}$ is not included in the analysis.

This new $Na_{2.97}Eu_{0.015}M^{III}P_3O_9N$ solid solution shows the counterintuitive relationship between their unitcell volume and photoluminescence properties similar to $M^I_{2.97}Eu_{0.015}AlP_3O_9N$. As plotted in **Figure 3c**, the solid solution between $Al^{3+}$ and $Ga^{3+}$ (*i.e.*, $M^{III}$=Al, Al$_{0.75}$Ga$_{0.25}$, Al$_{0.5}$Ga$_{0.5}$, Al$_{0.25}$Ga$_{0.75}$, Ga) reveals blue-shifted emission spectra ($M^{III}$=Al, $\lambda_{em}$=429 nm; $M^{III}$=Ga, $\lambda_{em}$=409 nm) with increasing $Ga^{3+}$ (**Figure 3d**). This shift seems like following the unit cell expansion of ~2.5 % with the incorporation of the larger $Ga^{3+}$ ion,



inducing a smaller CFS by enlarging the $Eu^{2+}$ substitution site volume. However, the inclusion of $In^{3+}$ does not lead to an additional shift in the emission spectra, despite the unit cell volume expanding another 2.2 %. A change in the bandwidth is also observed, with $M^{III}$=Ga having a narrower *fwhm* = 42 nm (2,493 cm$^{-1}$) compared to $M^{III}$ = Al having a *fwhm*=58 nm (3,217 cm$^{-1}$). The narrower emission when $M^{III}$ = Ga originates from the lack of [Na(3)O$_6$] site emission, supported by Gaussian fitting. The emission spectra are all identical for each excitation spectrum, unlike $M^{III}$=Al (**Figure 3e**). Exchanging $Ga^{3+}$ for $In^{3+}$ again does not change the peak shape.

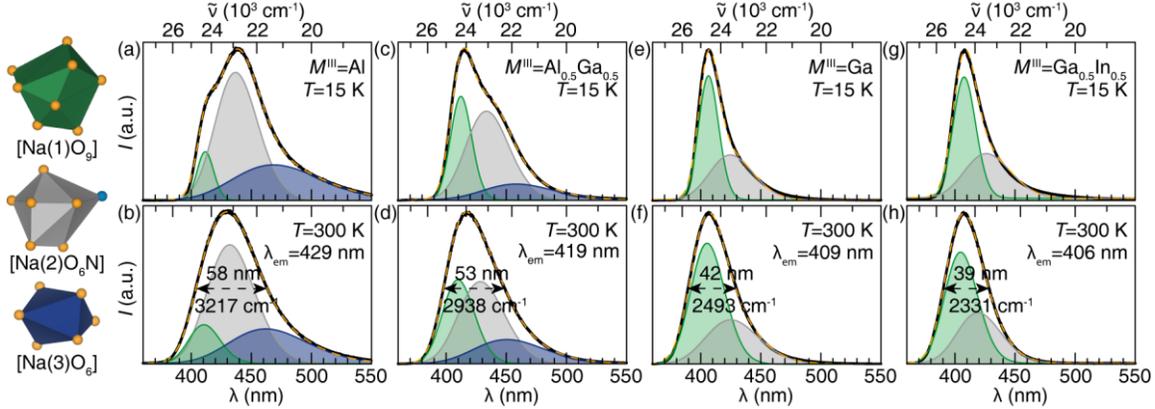

**Figure 4.** In the $Na_{2.97}Eu_{0.015}M^{III}P_3O_9N$ series, emission spectra are measured under 340 nm excitation at $T$=15 K and $T$=300K. Emission spectra of $M^{III}$=Al (a) at $T$=15 K, (b) at $T$=300 K, and $M^{III}$=Al$_{0.5}$Ga$_{0.5}$ (c) at $T$=15 K, (d) at $T$=300 K are deconvoluted in three peaks from each Na$^+$ sites. With the absence of Al$^{3+}$, the peak from [Na(3)O$_6$] is quenched in $M^{III}$=Ga (e) at $T$=15 K, (f) at $T$=300 K, and $M^{III}$=Ga$_{0.5}$In$_{0.5}$ (g) at $T$=15 K, (h) at $T$=300 K.

Collecting the emission spectra of $Na_{2.97}Eu_{0.015}M^{III}P_3O_9N$ [$M^{III}$=Al, (Al$_{0.75}$Ga$_{0.25}$), (Al$_{0.5}$Ga$_{0.5}$), (Al$_{0.25}$Ga$_{0.75}$), Ga, (Ga$_{0.75}$In$_{0.25}$), (Ga$_{0.5}$In$_{0.5}$)] at low temperature (15 K) and comparing these results to room temperature provides additional insights into the photoluminescent properties. Deconvoluting each emission demonstrates that the photoluminescence of the $M^{I}M^{III}P_3O_9N$ system is independent of the unit cell volume. The $M^{III}$=Al emission spectrum with Gaussian function at 15 K reveals a good fit with three curves (**Figure 4a**). Three peaks are also used to describe the data at 300 K, with the only differences being thermally induced spectral broadening and each peak position slightly blue-shifted, which can be explained by unit-cell thermal expansion (**Figure 4b**) and/or contribution from transitions from thermally populated higher-energy vibronic levels of the 5d$_1$ emitting level. In $M$ = Al$_{0.5}$Ga$_{0.5}$, the emission intensity at 15 K (**Figure 4c**) and 300 K (**Figure 4d**) increases for Eu$^{2+}$ on the [Na(1)O$_9$] site whereas the emission from Eu$^{2+}$ on the [Na(2)O$_6$N] and [Na(3)O$_6$] sites diminishes. When Al$^{3+}$ is fully exchanged for Ga$^{3+}$, the emission from Eu$^{2+}$ on the [Na(1)O$_6$] increases further, Eu$^{2+}$ on the [Na(2)O$_6$N] shows a greater decrease, and the emission from Eu$^{2+}$ on the [Na(3)O$_6$] is entirely lost even at low temperature. Meanwhile, the peak positions of Eu$^{2+}$ on [Na(1)O$_9$] and Eu$^{2+}$ on [Na(2)O$_6$N] are mostly preserved at around 410 nm and 430 nm, respectively. As a result of the strong preference for Eu$^{2+}$ to occupy the [Na(1)O$_9$] site, the total emission spectrum is noticeably blue-shifted and narrowed. It can be explained by the possibility that Ga$^{3+}$ induces more distortion on the smallest 6-coordinated cation site, making it too small to be substituted by Eu$^{2+}$ or the nearest neighbor impact. Interestingly, incrementing In$^{3+}$ into Ga$^{3+}$ generates no remarkable differences in the emission spectrum (**Figure 4g**). The emission of $M$=Ga$_{0.5}$In$_{0.5}$ at $T$=300 K also shows a narrowed spectrum with a slightly shifted [Na(2)O$_6$N] emission peak.

**3.3 Thermal and Chromatic Stability Changes in $Na_{2.97}Eu_{0.015}M^{III}P_3O_9N$**



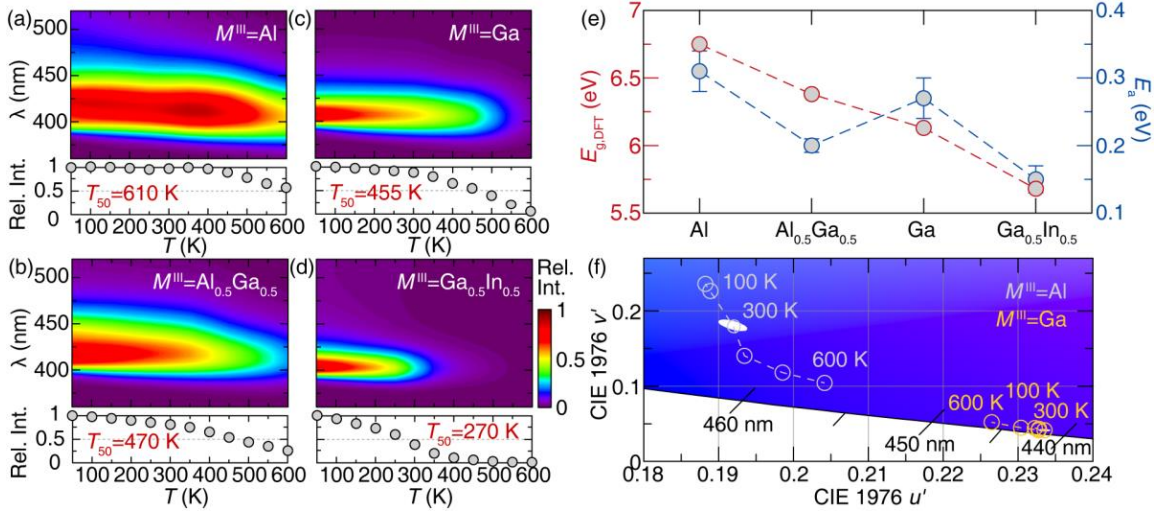

**Figure 5.** Temperature-dependent emission of $Na_{2.97}Eu_{0.015}M^{III}P_3O_9N$ series at (a) $M^{III}$=Al, (b) $M^{III}$=Al$_{0.5}$Ga$_{0.5}$, (c) $M^{III}$=Ga, and (d) $M^{III}$=Ga$_{0.5}$In$_{0.5}$. $T_{50}$ decreases along the diminishing host structure's bandgap. (e) Calculated $E_g$ using diffuse reflectance spectrum and $E_a$ values using temperature-dependent emission spectra of $Na_{2.97}Eu_{0.015}M^{III}P_3O_9N$ series (f) Changes in chromaticity by increasing temperature in CIE 1976 space of $M^{III}$=Al and $M^{III}$=Ga plotted with 3-step MacAdam ellipse at 300 K.

The operating temperature of a commercial LED package is approximately 423 K.[7] Thus, the emission intensity should be thermally robust against any thermal quenching processes at such a temperature. This is often qualitatively analyzed by the thermal quenching temperature ($T_{50}$), in which the relative emission intensity drops by 50%. Temperature-dependent emission measurements were conducted to study the series' thermal stability (**Figure 5a-d**). At $M^{III}$=Al, the $T_{50}$ shows an extrapolated value of 610 K, which is relatively high compared to other reported commercial thermally stable phosphors (YAG:Ce$^{3+}$, $T_{50}$≈632K; BAM:Eu$^{2+}$, $T_{50}$≈650K).[42,43] Although the $T_{50}$ decreases with an incremental Ga$^{3+}$ to 470 K for $M^{III}$=Al$_{0.5}$Ga$_{0.5}$, and to 455K for $M^{III}$=Ga, their values are still reasonably thermally stable, showing higher values than 423 K. Here, In$^{3+}$ dramatically reduces the photoluminescence thermal stability, decreasing the $T_{50}$ for $M^{III}$=Ga$_{0.5}$In$_{0.5}$ to 270 K with almost zero intensity at 423 K. To explain the reduced thermal stability, changes in the bandgap of hosts along their composition alteration should be examined first. The bandgap of $Na_3M^{III}P_3O_9N$ was calculated by DFT. Theoretically, $M^{III}$=Al should have the highest $E_g$ because of aluminum's smaller electronegativity (EN) than Ga$^{3+}$ or In$^{3+}$ ($\chi_{Al}$=1.61, $\chi_{Ga}$=1.81, $\chi_{In}$=1.78), which makes a more considerable EN difference between phosphorus, oxygen, and nitrogen. The calculated $E_{g,DFT}$ values are 6.75 eV for $M^{III}$=Al, 6.38 eV for $M^{III}$=Al$_{0.5}$Ga$_{0.5}$, 6.13 eV for $M^{III}$=Ga, and 5.68 eV for $M^{III}$=Ga$_{0.5}$In$_{0.5}$ (**Figure 5e**). Based on the bandgap values, the decreasing $T_{50}$ can be explained by the thermally activated photoionization, which occurs more readily with a smaller bandgap.[44]

Although the $E_g$ values nicely explain the thermal stability for $M^{III}$=Al, Al$_{0.5}$Ga$_{0.5}$, and Ga, further explanation is required for the low thermal stability of $M^{III}$=In. The position of the Eu$^{2+}$ 5$d$ orbital energy levels is splitting due to CFS, and once the electron on the most stabilized 5$d_1$ level is excited by the thermal energy, it can be transferred into the host's conduction band, creating non-radiative relaxation and diminishing the photoluminescence intensity.[45,46] The smaller bandgap decreases the energy level of the conduction band, enhancing photoionization quenching. The activated energy ($E_a$), which is the energy gap between the edge of the conduction band and the 5$d_1$ level of Eu$^{2+}$, determines the degree of the quenching process. From the temperature-dependent emission spectra, the $E_a$ values can be estimated by the Arrhenius



dependence, as depicted by **Equation 4**,

$$I(T) = \frac{I_0}{1+Ae^{(-E_a/kT)}} \tag{4}$$

where $I_0$ is the initial intensity, $E_a$ is the activation energy, $k$ is the Boltzmann constant, and $T$ is the temperature. For $M^{III}$=Al, $E_a$ is 0.31(3) eV, while $M^{III}$=Al$_{0.5}$Ga$_{0.5}$ and $M^{III}$=Ga gives lower values of 0.20(1) eV and 0.27(3) eV (**Figure S6**). The $E_a$ of $M^{III}$=Ga$_{0.5}$In$_{0.5}$ shows the lowest value of 0.15(2) eV, corresponding well with the thermal quenching tendency in the series. These estimated $E_a$ values support the lower thermal stability of the In$^{3+}$containing hosts since thermal ionization can occur more easily with smaller $E_a$.

Chromatic stability with the increasing temperature is also crucial in the actual application, as is the thermal stability. Since $M^{III}$=Ga shows less shift in peak position along the temperature due to the lack of [Na(3)O$_6$] emission, as shown in **Figure 4e**, the chromatic stabilities of Na$_{2.97}$Eu$_{0.015}$AlP$_3$O$_9$N and Na$_{2.97}$Eu$_{0.015}$AlP$_3$O$_9$N are calculated based on the temperature-dependent emission at each temperature. Coordinates of six different temperatures in the 100 – 600 K are plotted in CIE 1976 $u'v'$ space with a 3-step MacAdam ellipse, which is presented as a white oval centered at each phosphor's 300 K color coordinate (**Figure 5f**). Compared to $M^{III}$=Al, the shift in emission color of $M^{III}$=Ga is noticeably smaller for all the temperature ranges. To quantify the color changes, the concept of $\Delta u'\Delta v'$, the distance between the color coordinates at 300 K ($u'_{init.}$, $v'_{init.}$) and the average color coordinate at six different temperatures ($u'_{avg.}$, $v'_{avg.}$), is calculated by following **Equation 5**.[47]

$$\Delta u'\Delta v' = \sqrt{(u'_{avg.} - u'_{init.})^2 + (v'_{avg.} - v'_{init.})^2} \tag{5}$$

Na$_{2.97}$Eu$_{0.015}$AlP$_3$O$_9$N ($M^{III}$=Al) has the higher value of $\Delta u'\Delta v'$= 0.0119 than Na$_{2.97}$Eu$_{0.015}$GaP$_3$O$_9$N ($M$=Ga) whose value is $\Delta u'\Delta v'$= 0.0038, supporting $M^{III}$=Ga is more chromatically stable.

## 3.4 Quantum Yield Analysis through Thermoluminescence of Na$_{2.97}$Eu$_{0.015}$$M^{III}$P$_3$O$_9$N

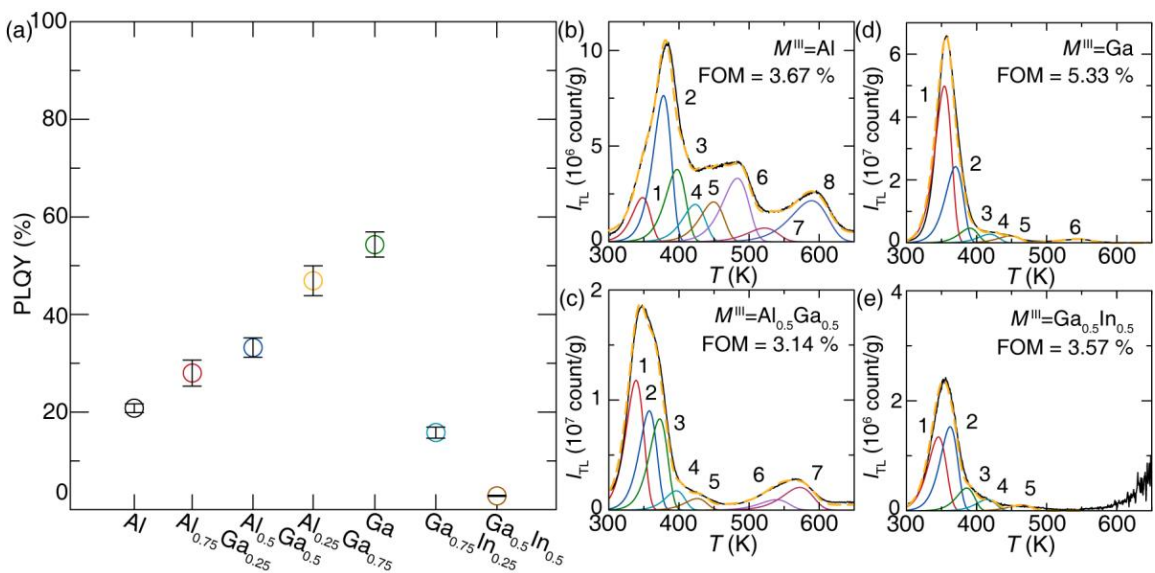

**Figure 6.** (a) Room-temperature PLQY of Na$_{2.97}$Eu$_{0.015}$$M^{III}$P$_3$O$_9$N. $M^{III}$=Ga shows the highest value which is almost three times as large as $M^{III}$=Al. The increment In$^{3+}$ significantly drops PLQY. TL glow curves of the solid solution at (b) $M^{III}$=Al, (c) $M^{III}$=Al$_{0.5}$Ga$_{0.5}$, (d) $M^{III}$=Ga, and (e) $M^{III}$= Ga$_{0.5}$In$_{0.5}$. Each curve is deconvoluted with the Randall-



Wilkins equation.

Even though the phosphor's thermal stability is satisfactory, the photoluminescent quantum yield should be higher for application. The most common commercial blue, BAM:$Eu^{2+}$, and green, $\beta$-SiAlON:$Eu^{2+}$, phosphors often show a PLQY > 90%. The PLQY of the $Na_{2.97}Eu_{0.015}M^{III}P_3O_9N$ solid solution was measured to investigate how the cation substitution in the host affects PLQY. Interestingly, PLQY increases from 21(2)% to 54(2)% between $Al^{3+}$ to $Ga^{3+}$ compositions, inversely related to the $E_g$ and $E_a$ tendencies (**Figure 6a**). However, the introduction of $In^{3+}$ rapidly quenches the PLQY to 2.2(4)%, indicating that the low quantum yield is related to the small $E_a$ and room-temperature thermal photoionization. The trend in quantum yield as a function of $M^{III}$ cation was also investigated in greater depth.

Due to the aliovalent substitution between $Na^+$ and $Eu^{2+}$, the inherent defect should occur to compensate for the charge valence. The most probable charge-neutral defect configurations are depicted in Kröger-Vink notation,

$$Na_{Na}^{\times} + Na_{Na}^{\times} \rightarrow Eu_{Na}^{\cdot} + V_{Na}' \tag{6}$$

$$Na_{Na}^{\times} + O_O^{\times} \rightarrow Eu_{Na}^{\cdot} + N_O' \tag{7}$$

where V refers to a vacancy site, subscript refers to the cation site, and superscript means the cation position, ×, ·, and ′ refer to neutral, positive, and negative charge, respectively.[48] Other common types of defects, like point defects and antisite defects, also can impact the $Na_{2.97}Eu_{0.015}M^{III}P_3O_9N$ series photoluminescence. To understand how the influence of defects along the cation substitution on $M^{III}$ differs in the host system, TL glow curves are analyzed (**Figure 6b-e**). The intensity of each TL curve was divided by a thermal quenching profile to correct the temperature dependency (TD) before being analyzed. Since the emission of $M^{III}$=$Ga_{0.5}In_{0.5}$ is mostly quenched after 400 K, the noise in its TL curve after 500 K is exaggerated by the TD correction; thus, the fitting was done before 500 K.

The peak positions and dosimetry parameters of trap states are determined using the glow curve deconvolution method. Deconvolution of each TL glow curve is performed with the GlowFit2 software using an approximation of the first-order kinetics Randall–Wilkins equation (**Equation 8**).[49,50]

$$I = n_0 s \exp\left(-\frac{E}{kT}\right) \exp\left(-\int_0^T \frac{s \exp\left(-\frac{E}{kT}\right)}{\beta} dT\right) \tag{8}$$

where $I$ is the TL intensity, $n_0$ is the number of trapped electrons, $s$ is the frequency factor, $E$ is trap depth, $k$ is Plank constant, $T$ is temperature, and $\beta$ is the heating rate. The area of each curve represents the bound electron population. The deconvolution parameters of each TL curve are plotted in **Table 1**. The values of $E$ and $s$ can evaluate the reliability of fitting. Additionally, since the $s$ reflects the interaction between the trap-bounded electron and the lattice phonon, the maximum value of $s$ should be the same or lower than the lattice vibration of the host. In the $Na_3M^{III}P_3O_9N$ ($M^{III}$=Al, Ga, In), the vibrations of O–N, O–P, N–P, and $M$–O are generally between 1,000 – 1,100 $cm^{-1}$, 900 – 1,000 $cm^{-1}$, 1,100 – 1,200 $cm^{-1}$, and 400 – 650 $cm^{-1}$, respectively, which correspond to the range from $1.2 \times 10^{13}$ to $3.6 \times 10^{13}$ $s^{-1}$.[51–54] The $s$ values extracted from the fitting are primarily between $1 \times 10^{11}$ and $3 \times 10^{12}$, supporting the validity of TL deconvolution.

**Table 1**. Parameters of calculated trap states using thermoluminescence glow curves.

| (a) $Na_{2.97}Eu_{0.015}AlP_3O_9N$ | | | | (b) $Na_{2.97}Eu_{0.015}Al_{0.5}Ga_{0.5}P_3O_9N$ | | | |
|---|---|---|---|---|---|---|---|
| Trap No. | $T$ (K) | $E$ (eV) | $s$ ($s^{-1}$) | Trap No. | $T$ (K) | $E$ (eV) | $s$ ($s^{-1}$) |
| 1 | 349.00 | 0.79 | $1.73 \times 10^{11}$ | 1 | 338.83 | 0.77 | $1.86 \times 10^{11}$ |



| | | | | | | | |
|---|---|---|---|---|---|---|---|
| 2 | 378.81 | 0.92 | $3.39 \times 10^{12}$ | 2 | 357.71 | 0.87 | $1.47 \times 10^{12}$ |
| 3 | 397.98 | 0.94 | $3.54 \times 10^{11}$ | 3 | 372.56 | 0.90 | $1.24 \times 10^{12}$ |
| 4 | 424.00 | 0.95 | $1.03 \times 10^{11}$ | 4 | 396.30 | 0.97 | $1.54 \times 10^{12}$ |
| 5 | 450.00 | 1.01 | $1.35 \times 10^{11}$ | 5 | 426.01 | 0.99 | $3.38 \times 10^{11}$ |
| 6 | 483.90 | 1.04 | $4.86 \times 10^{10}$ | 6 | 538.80 | 1.19 | $6.05 \times 10^{10}$ |
| 7 | 522.60 | 1.05 | $6.43 \times 10^{9}$ | 7 | 572.50 | 1.29 | $9.43 \times 10^{10}$ |
| 8 | 589.50 | 1.12 | $1.26 \times 10^{9}$ | | | | |

| (c) $Na_{2.97}Eu_{0.015}GaP_3O_9N$ | | | | (d) $Na_{2.97}Eu_{0.015}Ga_{0.5}In_{0.5}P_3O_9N$ | | | |
|---|---|---|---|---|---|---|---|
| Trap No. | $T$ (K) | $E$ (eV) | $s$ (s$^{-1}$) | Trap No. | $T$ (K) | $E$ (eV) | $s$ (s$^{-1}$) |
| 1 | 354.12 | 0.90 | $6.19 \times 10^{12}$ | 1 | 345.61 | 0.77 | $1.47 \times 10^{11}$ |
| 2 | 370.02 | 0.91 | $2.17 \times 10^{12}$ | 2 | 362.30 | 0.90 | $2.45 \times 10^{12}$ |
| 3 | 390.48 | 0.98 | $3.09 \times 10^{12}$ | 3 | 385.96 | 0.99 | $7.34 \times 10^{12}$ |
| 4 | 418.69 | 1.03 | $1.50 \times 10^{12}$ | 4 | 415.19 | 1.04 | $2.83 \times 10^{12}$ |
| 5 | 448.24 | 1.07 | $5.80 \times 10^{11}$ | 5 | 464.21 | 1.13 | $1.22 \times 10^{12}$ |
| 6 | 542.69 | 1.15 | $1.97 \times 10^{10}$ | | | | |

Notably, $M^{III}$=Al exhibits eight pronounced peaks (**Figure 6b**), and at least those numbered 3 – 8 are deep enough to immobilize the excited carriers for basically infinite time under ambient conditions. This phenomenon elucidates the low quantum yield observed in $M^{III}$=Al, as a significant portion of the excited electron population becomes trapped in various states, inhibiting photoluminescence despite the host material possessing the largest band gap and activation energy within the solid solution series. Along with the increasing $Ga^{3+}$, the number of trap states and the portion of higher energy trap states decrease noticeably, supporting the highest PLQY at $M^{III}$=Ga, for which no significant TL appears above 400 K. Incrementally increasing $In^{3+}$ does not show noticeable changes in trap states, indicating its low PLQY mainly stems from thermal quenching rather than defect quenching. Overall, $M^{III}$=Ga reveals the best overall thermoluminescence properties with the narrowest emission.

## 4. Conclusion

This series of $Eu^{2+}$-substituted $Na_3M^{III}P_3O_9N$ ($M^{III}$= Al, $Al_{0.75}Ga_{0.25}$, $Al_{0.5}Ga_{0.5}$, $Al_{0.25}Ga_{0.75}$, Ga, $Ga_{0.75}In_{0.25}$, $Ga_{0.5}In_{0.5}$) phosphors was successfully synthesized using gas reduction nitridation. Powder X-ray diffraction confirmed the quality of the products from the synthesis while photoluminescence measurement at low temperature supports that $Eu^{2+}$ replaces three crystallographically independent $Na^+$ sites in $Na_{2.97}Eu_{0.015}AlP_3O_9N$. Interestingly, incorporating $Ga^{3+}$ shifts the site selectivity so $Eu^{2+}$ prefers only two substitution sites, $[Na(1)O_9]$ and $[Na(2)O_6N]$. $Na_{2.97}Eu_{0.015}AlP_3O_9N$ generates 429 nm violet emission with $fwhm$=58 nm (3,217 cm$^{-1}$). As $Ga^{3+}$ increases, it appears there must be a distortion in the crystal structure, generating more blue-shifted emission further compounded by the lack of $Eu^{2+}$ on the $[Na(3)O_6]$ crystallographic site. This allows narrower emission with $fwhm$=42 nm (2,493 cm$^{-1}$) in $Na_{2.97}Eu_{0.015}GaP_3O_9N$ with $\lambda_{em}$=409 nm. The presence of $Ga^{3+}$ also provides a decent $T_{50}$=455 K, while $Na_{2.97}Eu_{0.015}GaP_3O_9N$ shows high chromatic stability. Additionally, PLQY of $Na_{2.97}Eu_{0.015}GaP_3O_9N$ is the highest among the series, with a reasonable value of 54(2) %, worthy of further optimization efforts. Thermoluminescence analysis supports that the highest quantum yield of $M^{III}$=Ga originates from a smaller



number of deep trap states that minimize their deleterious impact on excited electrons. This study of the Na$_{2.97}$Eu$_{0.015}$$M^{III}$P$_3$O$_9$N series offers that cation substitution on the host can accompany the preferred cation site selectivity, generating narrow and thermally and chromatically stable emission. This may be taken as proof that deliberately executed cation substitution may be an efficient approach for modeling and tuning important phosphors' properties, among them for designing narrow emitting and chromatically stable phosphors.

*ORCID*
Nakyung Lee: 0000-0001-5048-1816
Justyna Zeler: 0000-0002-2484-5484
Małgorzata Sójka: 0000-0001-9346-8929
Eugeniusz Zych: 0000-0001-5927-7942
Jakoah Brgoch: 0000-0002-1406-1352

*Notes*
The authors declare no competing financial interest.

## 5. Acknowledgments
The National Science Foundation (DMR-1847701) for supporting this work. J.Z. acknowledges support from the National Science Center (NCN) Poland project 2023/49/B/ST5/04265.

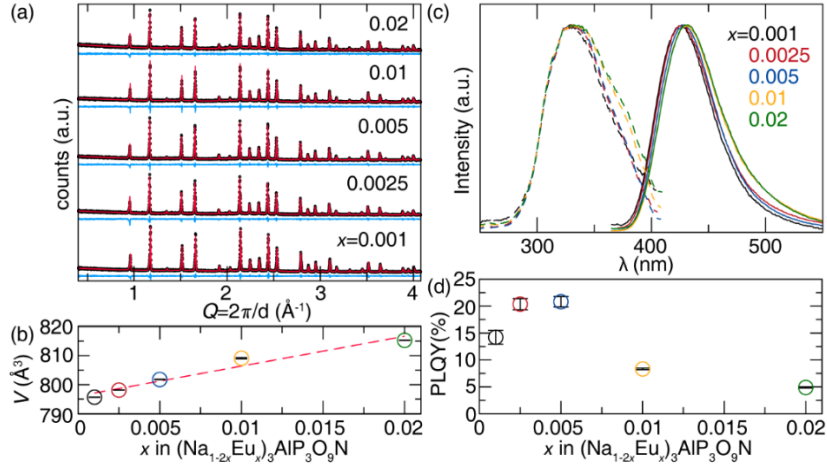

**Figure S1.** (a) Le Bail refinement of $(Na_{1-2x}Eu_x)_3AlP_3O_9N$ ($x$=0.001, 0.025, 0.005, 0.01, 0.02) with ICSD #78373 supports the purity of the series materials. (b) Calculated unit cell volumes from Le Bail refinements confirm the linear relation between $Eu^{2+}$ concentration and unit cell volume. (c) Excitation spectra are in dashed lines, and emission spectra are in solid lines of the $(Na_{1-x}Eu_x)_3AlP_3O_9N$ series. (d) Photoluminescent quantum yields of the series.

**Table S1.** Le Bail refinement data and statistics for $(Na_{1-2x}Eu_x)_3AlP_3O_9N$ solid solution.

| Composition | $x$=0.001 | $x$=0.0025 | $x$=0.005 | $x$=0.01 | $x$=0.02 |
|---|---|---|---|---|---|
| Radiation type, λ (Å) | Cu Kα, 1.5406 | | | | |
| 2θ range (deg) | 5 – 60 | | | | |
| Crystal System | Cubic | | | | |
| Space Group; Z | $P2_13$; 4 | | | | |
| a | 9.2664(4) | 9.2759(2) | 9.2901(3) | 9.3183(9) | 9.3419(3) |
| Volume (Å$^3$) | 795.66(1) | 798.12(3) | 801.80(8) | 809.1(2) | 815.28(6) |
| $_wR$ | 9.774 % | 9.844 % | 6.084 % | 9.205 % | 9.43 % |
| Reduced χ$^2$ | 1.71 | 1.66 | 1.97 | 1.71 | 2.79 |

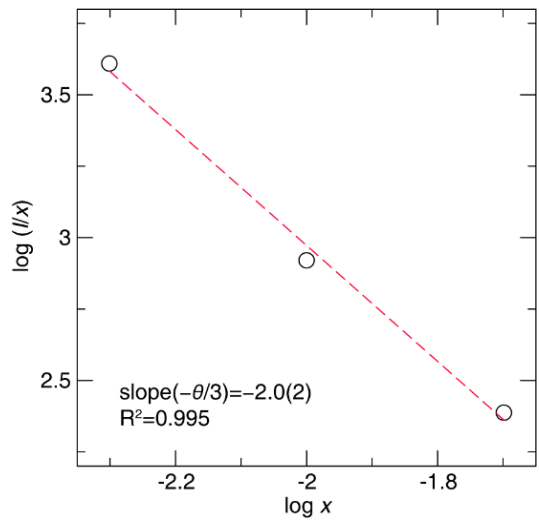

**Figure S2.** Log plot for the emission intensity under 340 nm excitation per the $Eu^{2+}$ concentration ($x$=0.005, 0.01, 0.02) in $(Na_{1-2x}Eu_x)_3AlP_3O_9N$ as a function of the $Eu^{2+}$ concentration.

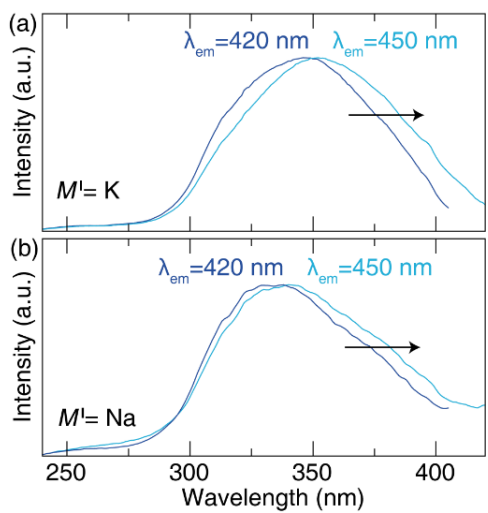

**Figure S3.** Excitation spectra of (a) $K_3AlP_3O_9N:Eu^{2+}$ and (b) $Na_{2.97}Eu_{0.015}AlP_3O_9N$ for different emission wavelengths.

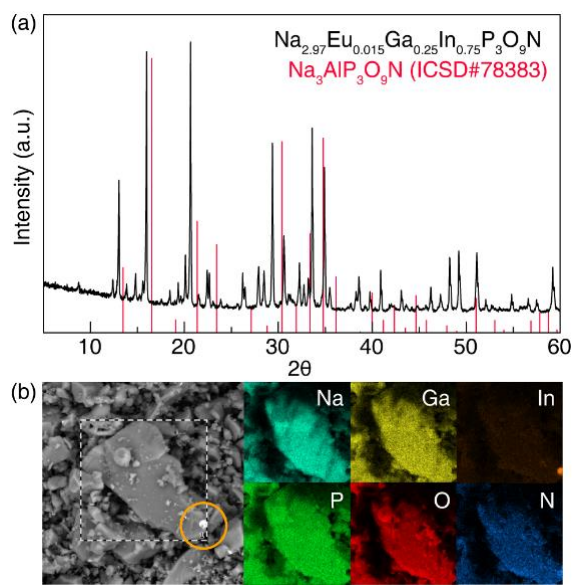

**Figure S4.** (a) Powder X-ray diffractogram of $Na_{2.97}Eu_{0.015}Ga_{0.25}In_{0.75}P_3O_9N$ with several impurity peaks. (b) SEM and EDS micrographs provide indium metal impurity in the material.

**Table S2.** Le Bail refinement data and statistics for $Na_{2.97}Eu_{0.015}M^{III}P_3O_9N$ solid solution.

| Composition | $M^{III}$=Al | $M^{III}$=Al$_{0.75}$Ga$_{0.25}$ | $M^{III}$=Al$_{0.5}$Ga$_{0.5}$ | $M^{III}$=Al$_{0.25}$Ga$_{0.75}$ |
|---|---|---|---|---|
| Radiation type, λ (Å) | Cu Kα, 1.5406 | | | |
| 2θ range (deg) | 5 - 60 | | | |
| Crystal System | Cubic | | | |
| Space Group; Z | $P2_13$; 4 | | | |
| a | 9.2901(3) | 9.3049(8) | 9.3189(1) | 9.3343(3) |
| Volume (Å$^3$) | 801.80(8) | 805.6(2) | 809.27(4) | 813.29(9) |
| $_wR$ | 6.084 % | 5.57 % | 5.628 % | 6.313 % |
| Reduced χ$^2$ | 1.97 | 1.79 | 1.85 | 2.12 |

| Composition | $M$=Ga | $M$=Ga$_{0.75}$In$_{0.25}$ | $M$=Ga$_{0.5}$In$_{0.5}$ | |
|---|---|---|---|---|
| Radiation type, λ (Å) | Cu Kα, 1.5406 | | | |
| 2θ range (deg) | 5 - 60 | | | |
| Crystal System | Cubic | | | |
| Space Group; Z | $P2_13$; 4 | | | |
| $a$ | 9.3637(1) | 9.4013(8) | 9.4346(4) | |
| Volume (Å$^3$) | 821.01(3) | 830.9(2) | 839.8(1) | |
| $_wR$ | 6.441 % | 11.983 % | 11.319 % | |
| Reduced $\chi^2$ | 2.17 | 2.40 | 2.07 | |

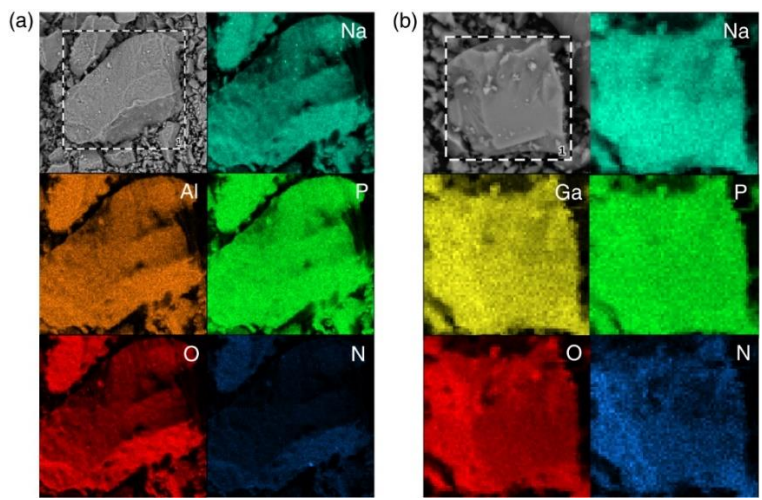

**Figure S5.** (a) SEM and EDX micrographs of Na$_{2.97}$Eu$_{0.015}$AlP$_3$O$_9$N and (b) Na$_{2.97}$Eu$_{0.015}$GaP$_3$O$_9$N. All elements in the composition except Eu$^{2+}$ with low concentration are uniformly distributed, showing no elemental separation.

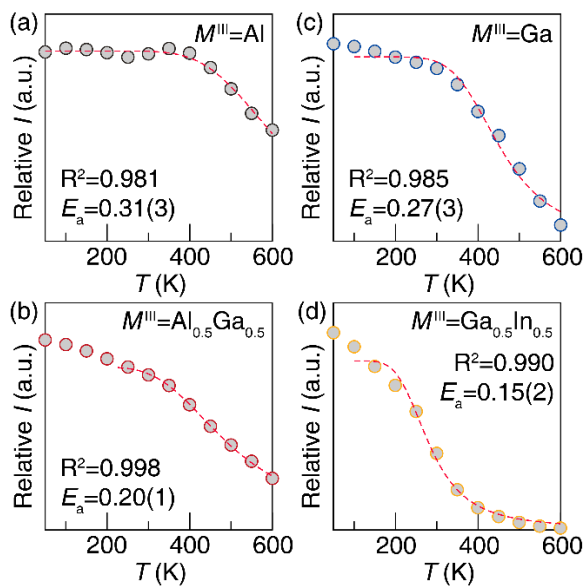

**Figure S6.** $E_a$ calculation calculation on the integrated intensity of temperature dependent emissions of the solid solution Na$_{2.97}$Eu$_{0.015}$$M^{III}$P$_3$O$_9$N at (a)$M^{III}$=Al, (b) $M^{III}$=Al$_{0.5}$Ga$_{0.5}$, (c) $M^{III}$=Ga, and (d) $M^{III}$= Ga$_{0.5}$In$_{0.5}$.